\documentclass[12pt]{article}
\usepackage{epsfig,graphicx}
\usepackage{amssymb}
\usepackage{cite}

\newcommand{\beq}{\begin{equation}}
\newcommand{\eeq}{\end{equation}}
\newcommand{\ba}{\begin{array}}
\newcommand{\ea}{\end{array}}
\newcommand{\beqa}{\begin{eqnarray}}
\newcommand{\eeqa}{\end{eqnarray}}
\newcommand{\lsim}{\stackrel{<}{_\sim}}
\newcommand{\gsim}{\stackrel{>}{_\sim}}

\newcommand{\cO}{{\cal O}}
\newcommand{\cB}{{\cal B}}
\newcommand{\BR}{{\cal B}}
\newcommand{\cL}{{\cal L}}

\newcommand{\no}{\nonumber}
\newcommand{\yuk}{{\lambda}}
\newcommand{\LLN}{{\Lambda_{\rm LN}}}
\newcommand{\LLFV}{{\Lambda}}
\newcommand{\Btaun}{{B \to \tau \nu}}
\newcommand{\msq}{M_{\tilde{q}}}
\newcommand{\msl}{M_{\tilde{\ell}}}

\def\npb#1#2#3{    {\it Nucl. Phys.}~B {\bf #1}, #3 (#2)}
\def\plb#1#2#3{    {\it Phys. Lett.}~B {\bf #1}, #3 (#2)}

\def\prd#1#2#3{    {\it Phys. Rev.}~D {\bf #1}, #3 (#2)}

\def\epjc#1#2#3{   {\it Eur. Phys. J.}~C {\bf #1}, #3 (#2)}

\def\Title#1{\begin{center} {\Large {\bf #1} } \end{center}}

\begin{document}

\Title{Shedding light on flavour symmetries \\
with rare decays of quarks and leptons }

\begin{center}{\large \bf Contribution to the proceedings of HQL06,\\
Munich, October 16th-20th 2006}\end{center}

\bigskip\bigskip

\begin{raggedright}  

{\it Gino Isidori\index{Isidori, Gino} \\
INFN, Laboratori Nazionali di Frascati,\\
Via E. Fermi 40, \\
I-00044 Frascati, Italy } 
\bigskip\bigskip
\end{raggedright}

%

\section{Introduction}

In the last few years there has been a great experimental 
progress in quark and lepton flavour physics. 
On the quark side, the two $B$-factories have 
been very successful, both from the accelerator and the 
detector point of view.  As a result, all the relevant parameters 
describing quark-flavour mixing within the Standard Model 
(quark masses and CKM angles) are now know with good accuracy. 
Despite this great progress, the overall picture of quark flavour physics 
is a bit frustrating as far as the search for physics beyond 
the Standard Model (SM) is concerned.
The situation is somehow similar to the flavour-conserving 
electroweak physics after LEP: the SM works very well and genuine 
one-loop electroweak effects have been tested with 
relative accuracy  in the $10\%$--$30\%$ range.

The situation of the lepton sector is more uncertain 
but also more exciting. The discovery of neutrino 
oscillations has two very significant implications:
i) the SM is not complete; ii) there exists new 
flavour structures in addition to the three 
SM Yukawa couplings. We have not yet enough information
to unambiguously determine how the SM Lagrangian 
should be modified in order to describe the 
phenomenon of neutrino oscillations. However, 
natural explanations point toward the 
existence of new degrees of freedom 
with explicit breaking of lepton number
at very high energy scales ($\Lambda_{\rm LN} \sim 10^{10}$--$10^{15}$~GeV), 
in agreement with the expectations of 
Grand Unified Theories (GUT). 
As I will discuss in this talk,  
these insight about non-SM degrees of freedom  
from neutrino physics are likely to have non-trivial 
implications in other sectors of the model. In particular, 
in rare decays of charged leptons and, possibly, 
in a few rare $B$ and $K$ decays. Interestingly enough,
these connections can be derived without specific 
dynamical assumptions about new physics, but 
only analysing the flavour-symmetry structure of the
theory be means of general Effective Field
Theory (EFT) approaches.

\section{The SM as EFT and the flavour problem}

The SM Lagrangian can be regarded as the renormalizable 
part of an effective field theory, valid up to some still undetermined 
cut-off scale $\Lambda$ above the electroweak scale. 
Since the SM is renormalizable, we have no direct
clues about the value of $\Lambda$; however, 
theoretical arguments based on a natural solution of the 
hierarchy problem suggest that $\Lambda$ should not exceed a few TeV.
As long as we are interested only in low-energy experiments, 
the EFT approach to physics beyond the SM is particularly 
useful. It allows us to analyse all realistic extensions 
of the model in terms of few unknown parameters (the 
coefficients of the higher-dimensional operators 
suppressed by inverse powers of $\Lambda$) and 
to compare the sensitivity to  New Physics (NP) 
of different low-energy observables.

The non-renormalizable 
operators should naturally induce large effects 
in processes which are not mediated by tree-level SM amplitudes, 
such as flavour-changing neutral-current (FCNC) rare processes. 
Up to now there is no evidence of deviations from the SM in these
processes and this 
implies severe bounds on the effective scale of various dimension-six
operators. For instance, the good agreement between SM 
expectations and experimental determinations of $K^0$--${\bar K}^0$ 
mixing leads to bounds above $10^4$~TeV for the effective scale 
of $\Delta S=2$ operators, i.e.~well above the few TeV 
range suggested by the Higgs sector. Similar bounds are 
obtained for the scale of operators 
contributing to lepton-flavour violating (LFV) 
transitions in the lepton sector, such as $\mu\to e\gamma$.

The apparent contradiction between these 
two determinations of  $\Lambda$ is a manifestation of what in 
many specific frameworks (supersymmetry, technicolour, etc.)
goes under the name of {\em flavour problem}:
if we insist with the theoretical prejudice that new physics has to 
emerge in the TeV region, we have to conclude that the new theory 
possesses a highly non-generic flavour structure. 
Interestingly enough, this structure has not been clearly identified yet,
mainly because the SM, i.e.~the low-energy 
limit of the new theory, doesn't possess an exact flavour symmetry.

The most reasonable (but also most {\em pessimistic}) solution
to the flavour problem is the so-called 
{\it Minimal Flavour Violation} (MFV) hypothesis \cite{Georgi,MFV2,MFV,MLFV}. 
Under this assumption, which will be discussed  
in detail in the next sections, flavour-violating 
interactions are linked to the
known structure of Yukawa couplings also beyond the SM. 
As a result, non-standard contributions in FCNC 
transitions turn out to suppressed to a level consistent 
with experiments even for $\Lambda \sim$~few TeV.
On the most interesting aspects of the MFV hypothesis 
is that it can easily be implemented within the 
general EFT approach to new physics \cite{MFV,MLFV}. 
The effective theories based on this symmetry principle
allow us to establish unambiguous correlations 
among NP effects in various rare decays.
These falsifiable predictions are the key ingredients   
to identify in a model-independent way which are the 
irreducible sources of breaking of the flavour symmetry.

\section{MFV in the quark sector}
The pure gauge sector of the SM is invariant under
a large symmetry group of flavour transformations: 
${\mathcal G}_{\rm SM} = {\mathcal G}_{q} \otimes 
{\mathcal G}_{\ell} \otimes U(1)^5$,
where 
\beq
{\mathcal G}_{q}
= {\rm SU}(3)_{Q_L}\otimes {\rm SU}(3)_{U_R} \otimes {\rm SU}(3)_{D_R},
\qquad 
{\mathcal G}_{\ell}
=  {\rm SU}(3)_{L_L} \otimes {\rm SU}(3)_{E_R}
\eeq
and three of the five $U(1)$ charges can be identified with 
baryon number, lepton number and hypercharge \cite{Georgi,MFV}. 
This large group and, particularly the ${\rm SU}(3)$ 
subgroups controlling flavour-changing transitions, is 
explicitly broken by the Yukawa interaction
\beq
\cL_Y  =   {\bar Q}_L \yuk_d D_R  H
+ {\bar Q}_L {\yuk_u} U_R  H_c
+ {\bar L}_L {\yuk_e} E_R  H {\rm ~+~h.c.}
\label{eq:LY}
\eeq
Since ${\mathcal G}_{\rm SM}$ is broken already within the SM, 
it would not be consistent to impose it as an exact symmetry 
of the additional degrees of freedom
present in SM extensions: even if absent a the tree-level,
the breaking of ${\mathcal G}_{\rm SM}$ would reappear at the quantum level 
because of the Yukawa interaction.  
The most restrictive hypothesis 
we can make to {\em protect} the breaking of ${\mathcal G}_{\rm SM}$ 
in a consistent way, is to assume that 
$\yuk_d$, $\yuk_u$ and $\yuk_e$ are the only source of 
${\mathcal G}_{\rm SM}$-breaking also beyond the SM.

To derive the phenomenological consequences of this hypothesis, 
it is convenient to treat ${\mathcal G}_{\rm SM}$ as an unbroken 
symmetry of the underlying theory, promoting  the $\yuk_i$ to be dynamical fields with 
non-trivial transformation properties under ${\mathcal G}_{\rm SM}$
\beq
\yuk_u \sim (3, \bar 3,1)_{{\rm SU}(3)^3_q}~,\qquad
\yuk_d \sim (3, 1, \bar 3)_{{\rm SU}(3)^3_q}~,\qquad 
\yuk_e \sim (3, \bar 3)_{{\rm SU}(3)^2_\ell}~.
\eeq
If the breaking of ${\mathcal G}_{\rm SM}$ occurs at very high energy scales 
 --well above the TeV region where the we expect new degrees of freedom--
at low-energies we would only be sensitive to the background values of 
the $\yuk_i$, i.e. to the ordinary SM Yukawa couplings. 
Employing the EFT language, 
we then define that an effective theory satisfies the criterion of
Minimal Flavour Violation if all higher-dimensional operators,
constructed from SM and $\yuk$ fields, are (formally)
invariant under the flavour group ${\mathcal G}_{\rm SM}$ \cite{MFV}. 

According to this criterion, one should in principle 
consider operators with arbitrary powers of the (adimensional) 
Yukawa fields. However, a strong simplification arises by the 
observation that all the eigenvalues of the Yukawa matrices 
are small, but for the top one, and that the off-diagonal 
elements of the CKM matrix ($V_{ij}$) are very suppressed. 
It is then easy to realize that, similarly to the pure SM case, 
the leading coupling ruling all FCNC transitions 
with external down-type quarks is \cite{MFV}:
\beq
(\Delta^q_{\rm LL})_{i\not=j} =  (\yuk_u \yuk_u^\dagger)_{ij}
\approx y_t^2  (V_{\rm CKM})^*_{3i} (V_{\rm CKM})_{3j}~,
\qquad 
  y_t =m_t/v \approx 1~.
\eeq 
As a result, within this framework the bounds on 
the scale of dimension-six FCNC effective operators 
turn out to be in the few TeV range (see Ref.~\cite{UTfits}
for updated values). Moreover, the flavour structure 
of $\Delta^q_{\rm FC}$ implies a well-defined link among 
possible deviations from the SM in FCNC transitions 
of the type $s\to d$, $b\to d$, and  
$b\to s$ (the only quark-level transitions where 
observable deviations from 
the SM are expected).

\begin{figure}[t]
\begin{center}
\includegraphics[width=55mm]{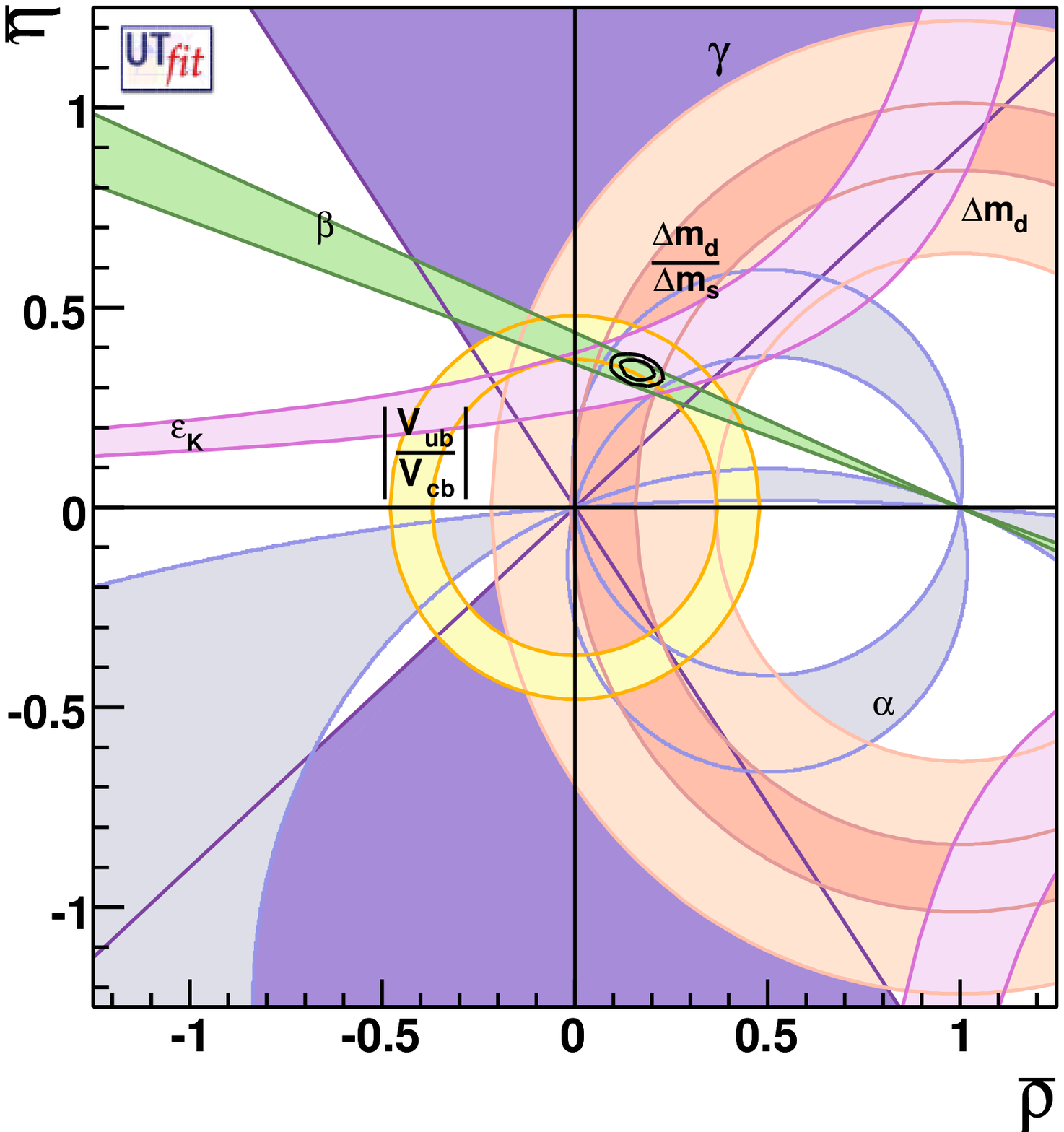}
\includegraphics[width=55mm]{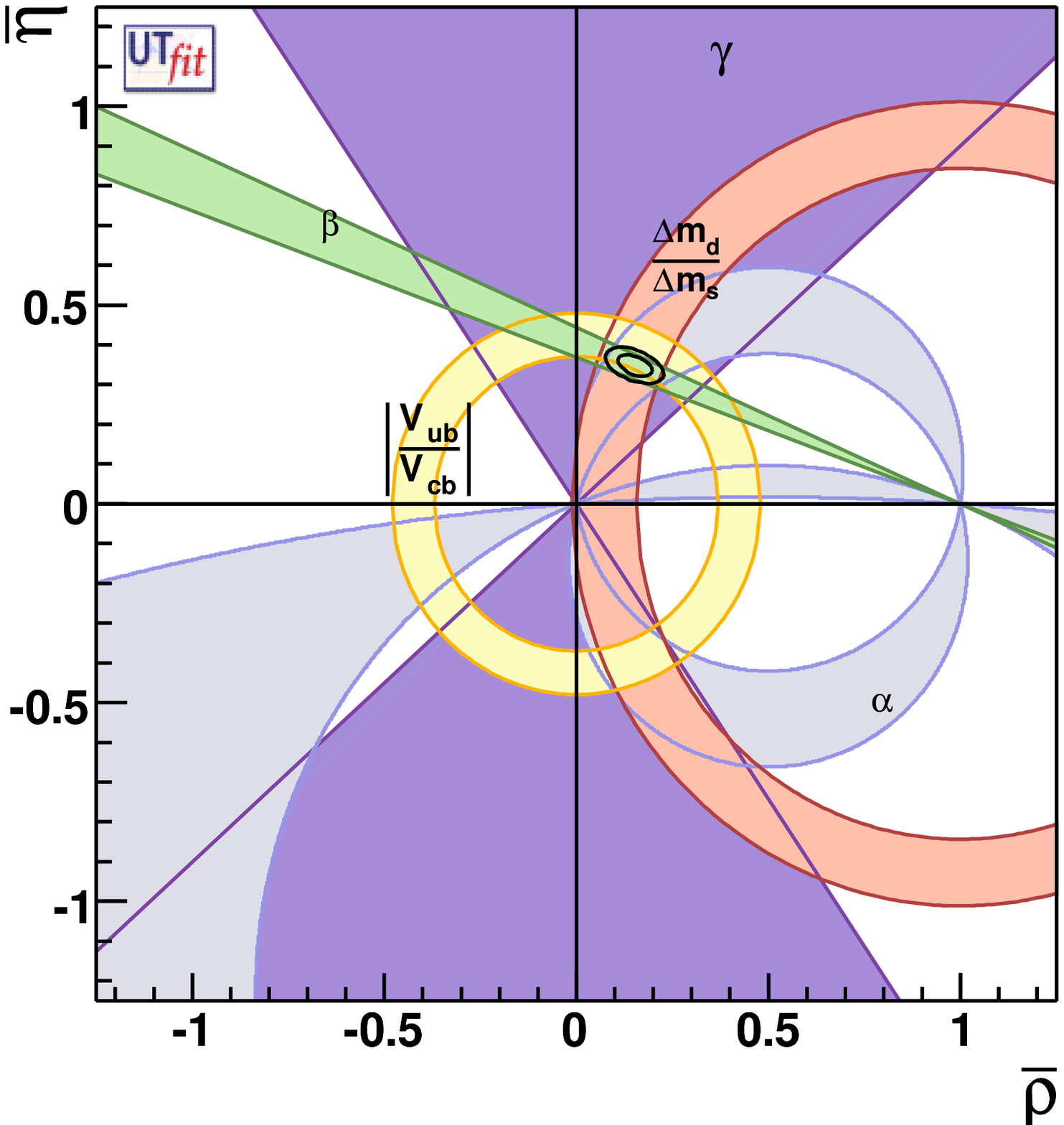}
\caption{\label{fig:UTfits} Fit of the CKM unitarity triangle within the SM (left) and 
in generic extensions of the SM satisfying the MFV hypothesis (right)~\cite{UTfits}. }
\end{center}
\end{figure}

The idea that the CKM matrix rules the strength of FCNC 
transitions also beyond the SM has become a very popular 
concept in the recent literature and has been implemented 
and discussed in several works (see e.g.~Ref.~\cite{MFV2}). 
However, it is worth stressing that the CKM matrix 
represents only one part of the problem: a key role in
determining the structure of FCNCs is also played  by quark masses
(via the GIM mechanism), or by the Yukawa eigenvalues. 
In this respect, the above MFV criterion provides the maximal protection 
of FCNCs (or the minimal violation of flavour symmetry), since the full 
structure of Yukawa matrices is preserved. Moreover, 
contrary to other approaches, the above MFV criterion 
is based on a renormalization-group-invariant symmetry argument,
which can easily be extended to EFT approaches where new degrees
of freedoms (such as extra Higgs doubles, or SUSY partners of the SM
fields) are explicitly included.

As shown in Fig.~\ref{fig:UTfits}, the MFV hypothesis provides a
natural (a posteriori) justification of why no NP effects have 
been observed in the quark sector: by construction, most of the clean 
observables measured at $B$ factories are insensitive to NP effects 
in this framework. However, it should be stressed that we are still
very far from having proved the validity of this hypothesis from data.
Non-minimal sources of flavour symmetry breaking with specific 
flavour structures, such as those discussed in Ref.~\cite{Mannel}, are  
still allowed (even with NP scales in the TeV range).  
A proof of the MFV hypothesis 
can be achieved only with a positive evidence of physics beyond 
the SM exhibiting the flavour pattern (link between $s\to d$, $b\to d$, and  
$b\to s$) predicted by the MFV assumption.

\section{MFV in the lepton sector}
Apart from arguments based on the analogy between quarks
and leptons, the introduction of a MFV hypothesis 
for the lepton sector (MLFV) is demanded 
by a severe fine-tuning problem also in the lepton sector:
within a generic EFT approach, the non-observation of 
$\mu\rightarrow e\gamma$ implies an effective NP scale 
above $10^5$ TeV unless the coupling of the corresponding 
effective operator is suppressed by some symmetry principle. 

Since the observed neutrino mass parameters are not 
described by the SM Yukawa interaction in Eq.~(\ref{eq:LY}),
the formulation of a MLFV hypothesis is not straightforward.
A proposal based on the assumption that the breaking 
of total lepton number (LN) and lepton flavour are decoupled in the underlying theory
has recently been presented in Ref.~\cite{MLFV},
and further analysed in Ref.~\cite{CG}.  
Two independent  MLFV scenarios have been identified. 
They are characterized by the different status assigned to
the effective Majorana mass matrix $g_\nu$  appearing as coefficient of 
the $|\Delta L| = 2$ dimension-five operator in the low 
energy effective theory~\cite{Weinberg:1979sa}:
\beq
\cL_{\rm eff}^{\nu}  = -\frac{1}{ \Lambda_{\rm LN}}\,g_\nu^{ij}(\bar
L^{ci}_L\tau_2 H)(H^T\tau_2L^j_L)  {\rm ~+~h.c.} \quad \longrightarrow 
\quad m_\nu =  \frac{g_\nu v^2}{\Lambda_{\rm LN}}
\eeq
In the truly minimal scenario (dubbed {\em minimal field content}),  
$g_\nu$ and the charged-lepton
Yukawa coupling ($\lambda_e$) are assumed to be the only irreducible sources 
of breaking of ${\mathcal G}_{\ell}$, the lepton-flavour 
symmetry of the low-energy theory. 

The irreducible character of $g_\nu$
does not hold in many realistic underlying 
theories with heavy right-handed neutrinos. For this reason, 
a second scenario (dubbed {\em extended field content}), 
with heavy right-handed neutrinos and a larger 
lepton-flavour symmetry group, ${\mathcal G}_{\ell}\times {\rm O}(3)_{\nu_R}$,
has also been considered. 
In this extended scenario, the most natural and 
economical choice about the symmetry-breaking terms is the identification 
of the two Yukawa couplings, $\lambda_\nu$ and $\lambda_e$, 
as the only irreducible symmetry-breaking structures.
In this context,
$g_\nu \sim  \lambda_\nu^T \lambda_\nu$ and the 
LN-breaking  mass term of the heavy 
right-handed neutrinos is flavour-blind 
(up to Yukawa-induced corrections):
\beqa
\cL_{\rm heavy}  &=&  -\frac{1}{2} M_\nu^{ij}\bar \nu^{ci}_R\nu_R^j {\rm ~+~h.c.} 
\qquad M_\nu^{ij}=M_\nu  \delta^{ij}  \no\\
\cL^{\rm ext}_{Y}  &=&  \cL_{Y}  +i \lambda_\nu^{ij}\bar\nu_R^i(H^T \tau_2L^j_L) {\rm ~+~h.c.} 
\eeqa
In this scenario the flavour changing coupling 
relevant to $l_i\rightarrow l_j\gamma$ decays reads
\begin{eqnarray}
\label{eq:MVFDRLext}
(\Delta^{\ell}_{\rm LR} )_{\rm MLFV} ~ \propto ~ \lambda_e \lambda_\nu^\dagger
\lambda_\nu 
 \to  \frac{m_e}{v}\, \frac{M_\nu}{v^2}U_{\rm PMNS}\, (m^{1/2}_{\nu})_{\rm diag} H^2
(m^{1/2}_{\nu})_{\rm diag}\,
U^\dagger_{\rm PMNS}
\end{eqnarray} 
where  $H$ is an Hermitian-orthogonal matrix  which can
be parametrized in terms of three real parameters ($\phi_i$) which
control the amount of CP-violation in the right-handed sector~\cite{Pascoli:2003rq}.
In the CP-conserving limit, $H \rightarrow I$ and the 
phenomenological predictions for lepton FCNC decays 
turns out to be quite similar to the minimal
field content scenario \cite{MLFV}. 

Once the field content of model is extended, there are in principle many 
alternative options to define the irreducible sources of lepton flavour 
symmetry breaking (see e.g.~Ref.\cite{Davidson:2006bd} for an extensive discussion). 
However, the specific choice discussed above has two important advantages:
it is predictive and closely resemble the MFV hypothesis in the quark sector. 
The $\nu_R$'s are the counterpart of right-handed up quarks and, similarly to 
the quark sector, the symmetry-breaking sources are two Yukawa couplings.

The basic assumptions of the MLFV hypotheses 
are definitely less data-driven 
with respect to the quark sector. Nonetheless, the formulation of 
an EFT based on these assumptions is still very useful.
As I will briefly illustrate in the following, it allows us to 
address in a very general way the following fundamental 
question: how can we detect the presence of new irreducible 
(fundamental) sources of LF symmetry breaking? 

\subsection{Phenomenological consequences on LFV decays}

Using the MLFV-EFT approach, one can easily 
demonstrate that --in absence of new sources of LF violation--
visible FCNC decays of $\mu$ and $\tau$ can occur only 
if there is a large hierarchy between $\LLFV$
(the scale of  new degrees of freedoms carrying LF) 
and $\LLN \sim M_\nu$ (the scale of total LN violation) \cite{MLFV}.
This condition is indeed realized within the explicit 
extensions of the SM widely discussed in the literature 
which predict sizable LF violating effects in charged 
leptons (see e.g.~Ref.~\cite{Barbieri,hisano,profumo,Herrero}).

\begin{figure}[t]
\begin{center}
\includegraphics[width=150mm]{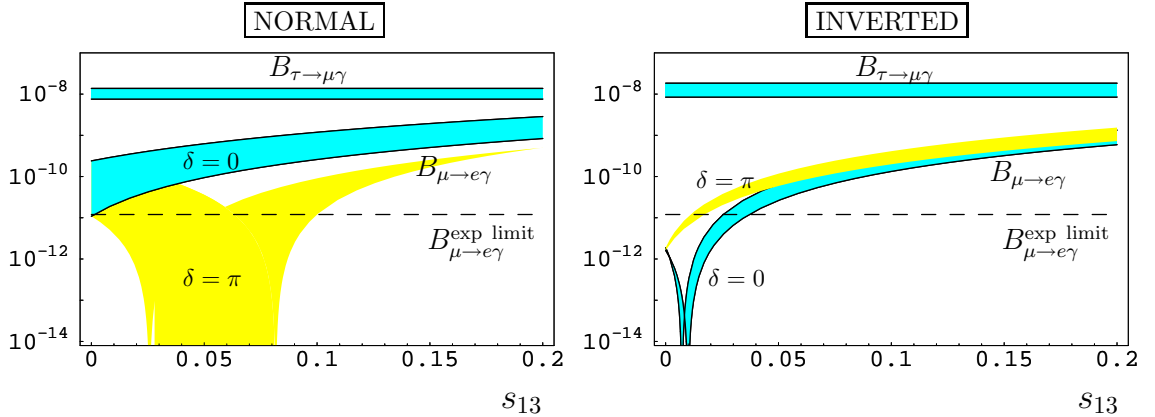}
\caption{\label{fig:MLFV}  
$B_{\tau \to  \mu \gamma} = \Gamma(\tau \to \mu \gamma)/\Gamma(\tau \to \mu \nu \bar{\nu}) $
compared to the $\mu  \to  e \gamma$ constraint within MLFV (minimal field
content),  as a function of the neutrino mixing angle $s_{13}$ \cite{MLFV}.
The shading corresponds to different values of the phase $\delta$ 
and the normal/inverted spectrum. The NP scales have been set to
 $\Lambda_{\rm LN}/\Lambda = 10^{10}$; their variation affects 
only the overall vertical scale.}
\end{center}
\end{figure}

\begin{figure}[t]
\begin{center}
\includegraphics[width=73mm,height=50mm]{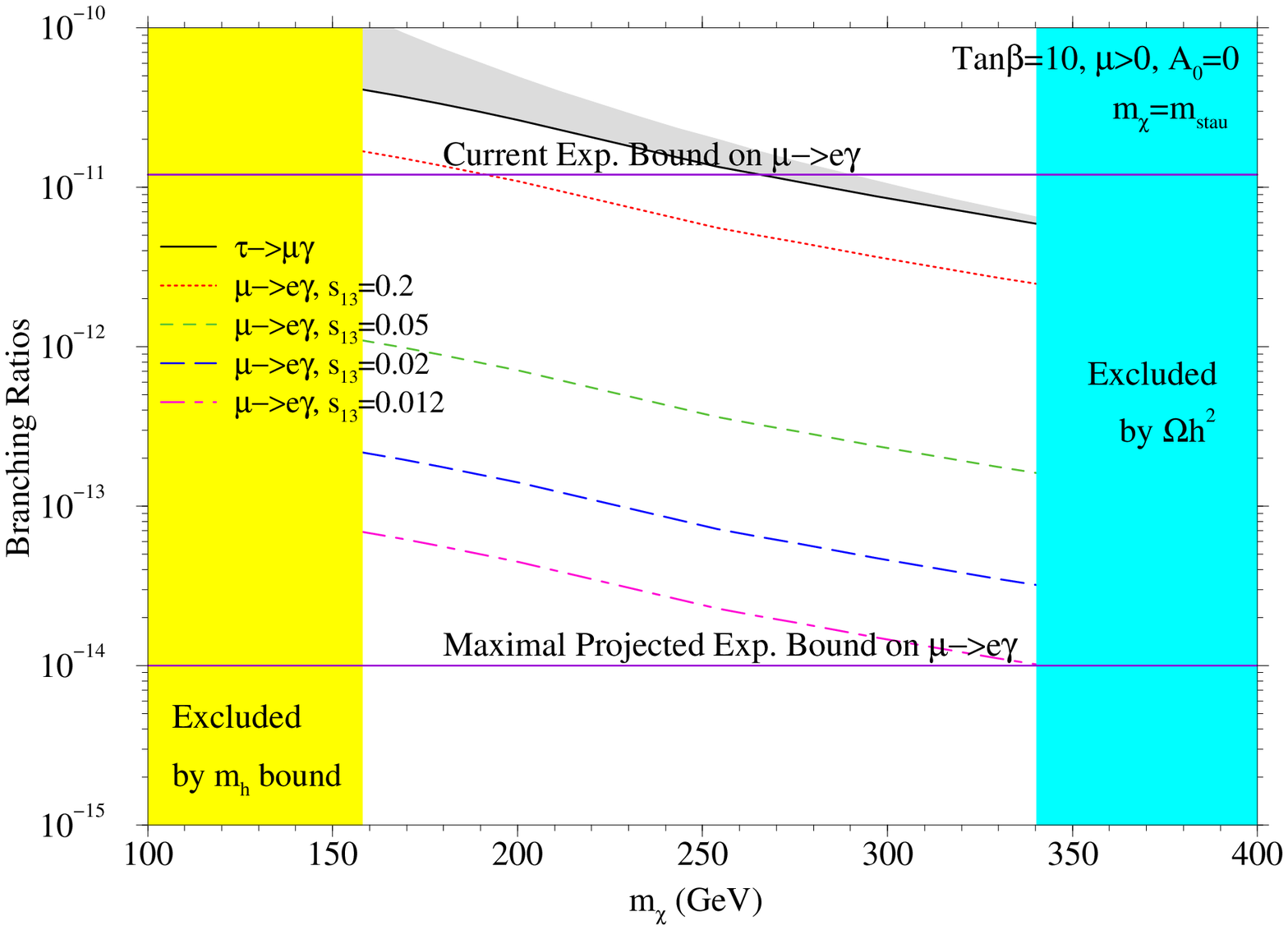}
\hskip 0.1 cm 
\includegraphics[width=73mm,height=50mm]{herrero.eps}
\caption{\label{fig:profumo}  
Left: Isolevel curves for $\cB(\mu\rightarrow e\gamma)$ and 
$\cB(\tau\rightarrow \mu\gamma)$ in the MSSM scenario of  Ref.~\cite{profumo}.
Right: $\cB(\mu\rightarrow e\gamma)$ vs. $\cB(\tau\rightarrow \mu\gamma)$ in the 
MSSM scenario of Ref.~\cite{Herrero}  }
\end{center}
\end{figure}

More interestingly, the EFT  allows us to draw unambiguous 
predictions about the relative size of LF violating decays 
of charged leptons (in terms of neutrino masses and mixing angles). 
At present, the uncertainty in the predictions for such ratios is limited 
from the poorly constrained value of the $1$--$3$ mixing angle in the neutrino
mass matrix ($s_{13}$) and, to a lesser extent, 
from the neutrino spectrum ordering and the CP violating phase $\delta$. 
One of the  clearest consequences from the phenomenological
point of view is the observation that  if $s_{13} \gsim 0.1$ there 
is no hope to observe $\tau \to \mu \gamma$ at future accelerators 
(see Fig.~\ref{fig:MLFV}). This happens because the stringent
constraints from $\mu \to e \gamma$ already forbid too low values 
for the effective scale of LF violation. In other words,
in absence of new sources of LF violation the most sensitive 
FCNC probe in the lepton sector is $\mu \to e \gamma$.
This process should indeed be observed at  MEG \cite{MEG}
for very realistic values of the new-physics scales $\LLFV$ 
and $\LLN$.  Interestingly enough, this conclusion holds 
both in the minimal- and in the extended-field-content
formulation of the MLFV framework. 

The expectation of a higher NP sensitivity of 
$\mu \to \mu \gamma$ with respect to 
$\tau \to \mu \gamma$
(taking into account the corresponding experimental resolutions) 
is confirmed in several realistic NP frameworks.
This happens for instance in the MSSM scenarios 
analysed in Ref.~\cite{hisano,profumo,Herrero} (see Fig.~\ref{fig:profumo})
with the exception of specific corners 
of the parameter space~\cite{hisano}.

\subsection{Leptogenesis}\label{sec:MFVlep}

In the MLFV scenario with extended field content we can hope 
to generate the observed matter-antimatter asymmetry of the 
Universe by means of leptogenesis~\cite{Yanagida}. The viability 
of leptogenesis within the MLFV framework, which has recently been
demonstrated in Ref.~\cite{Cirigliano:2006nu,Branco:2006hz,Uhlig:2006xf},
is an interesting conceptual point: it implies that there are no 
phenomenological motivations to introduce new sources of flavour symmetry 
breaking in addition to the four $\lambda_i$ (the three SM Yukawa couplings 
and  $\lambda_\nu$).  

\begin{figure}[t]
\begin{center}
\includegraphics[width=73mm,height=50mm]{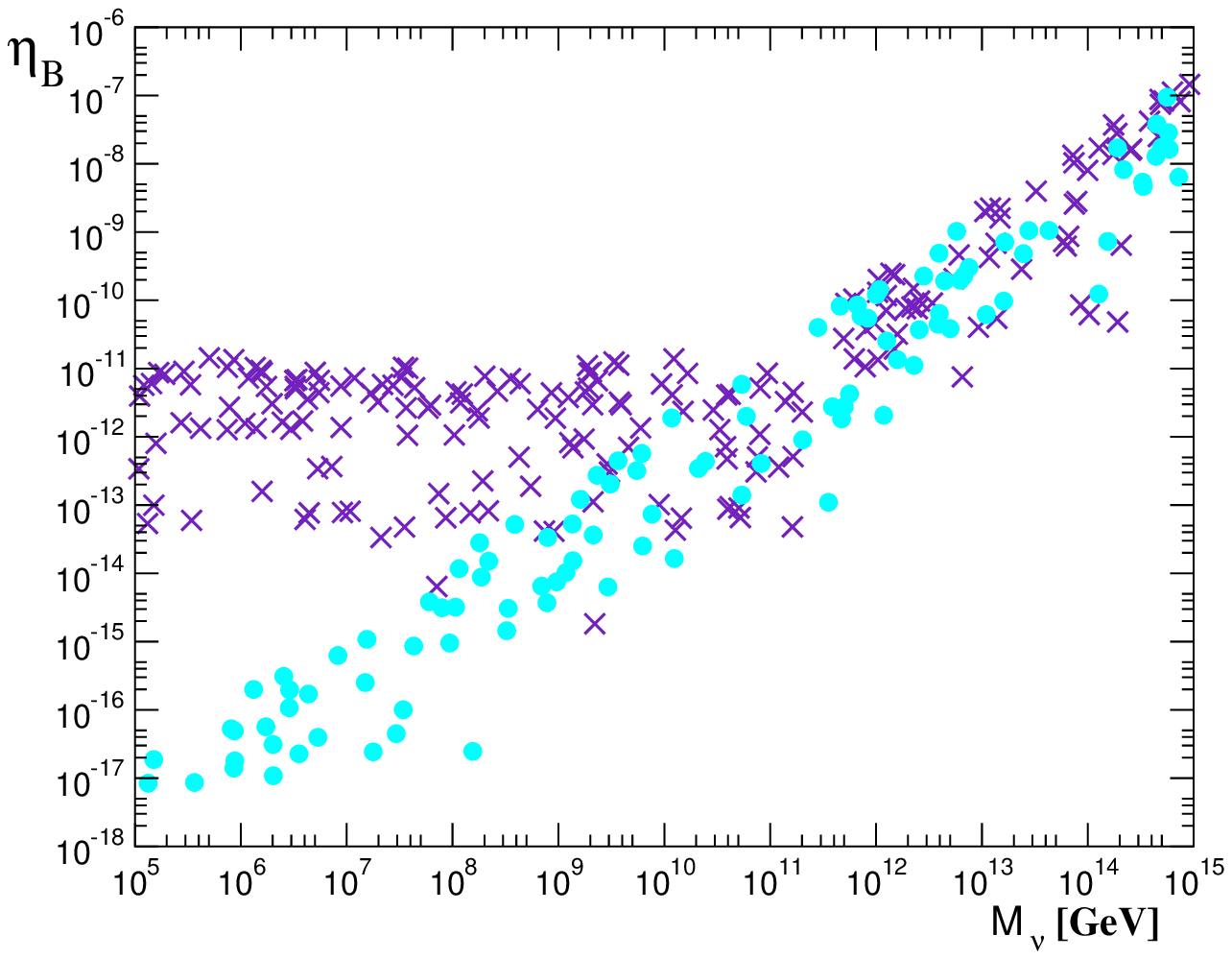}
\hskip 0.1 cm
\includegraphics[width=73mm,height=50mm]{Buras_lepto.eps}
\caption{Left: Baryon asymmetry ($\eta_B$) as a function of the right-handed
neutrino mass scale ($M_\nu$) for $c_{\nu l} = 0$ (cyan circles) and
$c_{\nu l} \neq 0$ (violet crosses)  
in the MLFV framework with extended field content~\cite{Cirigliano:2006nu}.
Right: $\eta_B$ as a function of $M_\nu$ with the inclusion of 
flavour-dependent effects \cite{Branco:2006hz}.
}\label{fig:MFVeta}
\end{center}
\end{figure}

A necessary condition for leptogenesis to occur is a 
non-degenerate heavy-neutrino spectrum. Within the MLFV
framework, the tree-level degeneracy of heavy neutrinos 
is lifted only by radiative corrections, which implies a rather
predictive/con\-stra\-ined scenario. The most general form of the  
$\nu_R$ mass-splittings has the following structure:
\begin{eqnarray} 
&&
\frac{\Delta M_R}{M_R} = c_\nu
\left[ \lambda_\nu \lambda_\nu^\dagger + (\lambda_\nu
\lambda_\nu^\dagger)^T  \right] \nonumber + c^{(1)}_{\nu\nu} \left[
\lambda_\nu \lambda_\nu^\dagger \lambda_\nu \lambda_\nu^\dagger +
(\lambda_\nu \lambda_\nu^\dagger \lambda_\nu \lambda_\nu^\dagger )^T
\right] \nonumber  \\
&&\quad  +c^{(2)}_{\nu\nu} \left[ \lambda_\nu
\lambda_\nu^\dagger (\lambda_\nu \lambda_\nu^\dagger)^T \right]
+ c^{(3)}_{\nu\nu}  \left[ (\lambda_\nu \lambda_\nu^\dagger)^T
\lambda_\nu \lambda_\nu^\dagger \right]\nonumber  + c_{\nu l} \left[
\lambda_\nu \lambda_e^\dagger \lambda_e \lambda_\nu^\dagger +
(\lambda_\nu \lambda_e^\dagger \lambda_e \lambda_\nu^\dagger )^T
\right] + \ldots
\end{eqnarray}
Even without specifying the value of the $c_i$, this form allows us 
to derive a few general conclusions \cite{Cirigliano:2006nu}:
\begin{itemize}
\item
The term proportional to $c_\nu$ does not generate 
a CPV asymmetry, but sets the scale for  
the mass splittings: these are of the order of 
magnitude of the decay widths, realizing in a natural way the 
condition of resonant leptogenesis. 
\item
The right amount of leptogenesis can be generated even 
with $\lambda_e = 0$, if all the $\phi_{i}$ (the CP-violating parameters of $H$)
are non vanishing.  However, since $\lambda_\nu\sim \sqrt{M_\nu}$, for low values of $M_\nu$ 
($\lesssim 10^{12}$ GeV) the asymmetry generated by the $c_{\nu l}$ term
dominates. In this case $\eta_B$ is typically too small to match the observed 
value and has a flat dependence on $M_\nu$. At $M_\nu \gtrsim 10^{12}$ GeV 
the quadratic terms $c^{(i)}_{\nu\nu}$ dominate, determining an 
approximate linear growth of  $\eta_B$ with $M_\nu$. These two regimes 
are illustrated in Fig.\ref{fig:MFVeta}.
\end{itemize}

\noindent
As shown in in Fig.\ref{fig:MFVeta},
baryogenesis through leptogenesis is viable in MLFV models.
In particular, assuming a loop hierarchy between the $c_i$ 
(as expected in a perturbative scenario) and neglecting 
flavour-dependent effects in  the Boltzmann equations 
(one-flavour approximation of Ref.\cite{Blanchet:2006dq}), 
the right size of $\eta_B$ is naturally reached for 
$M_\nu \gtrsim 10^{12}$~GeV~\cite{Cirigliano:2006nu}. 
As shown in Ref.~\cite{Branco:2006hz,Uhlig:2006xf}, 
this lower bound can be weakened by the inclusion of flavour-dependent  
effects in the Boltzmann equations and/or by the $\tan\beta$-enhancement of 
$\lambda_e$ occurring in two-Higgs doublet models.

From the phenomenological point of view, an important difference
with respect to the CP-con\-ser\-ving case is the fact that 
non-vanishing $\phi_i$ change the predictions of the LFV decays, 
typically producing an 
enhancement of the  $\mathcal{B}(\mu\to e\gamma)/\mathcal{B}(\tau\to \mu\gamma)$
ratio. The effect of the new phases 
is moderate and the CP-conserving predictions 
are recovered only for $M_\nu\gg 10^{12}$ GeV.

\section{MFV in Grand Unified Theories}
\label{sec:MFVgut}
Once we accept the idea that flavour dynamics obeys a MFV
principle, both in the quark and in the lepton sector, it is
interesting to ask if and how this is compatible with a
grand-unified theory (GUT), where quarks and leptons sit in the same
representations of a unified gauge group. This question has 
recently been addressed in Ref.~\cite{Grinstein:2006cg}, 
considering the exemplifying case of ${\rm SU}(5)_{\rm gauge}$.

Within ${\rm SU}(5)_{\rm gauge}$, the down-type singlet quarks ($d^c_{iR}$) and the lepton doublets 
($L_{iL}$) belong to the $\bar {\bf 5}$ representation; the quark doublet
($Q_{iL}$), the up-type ($u^c_{iR}$) and lepton singlets ($e_{iR}^c$) 
belong to the ${\bf 10}$ representation, and finally 
the right-handed neutrinos ($\nu_{iR}$) are singlet.
In this framework the largest 
group of flavour transformation commuting with the gauge group is
${\mathcal G}_{\rm GUT} = {\rm SU}(3)_{\bar 5} \times {\rm SU}(3)_{10}\times {\rm SU}(3)_1$, 
which is smaller than the direct product 
of the quark and lepton groups discussed before 
(${\mathcal G}_q \times {\mathcal G}_l$).  
We should therefore expect some violations of the MFV+MLFV predictions,
either in the quark sector, or in the lepton sector, or in both. 

A phenomenologically acceptable description of the low-energy fermion 
mass matrices requires the introduction of at least four irreducible 
sources of ${\mathcal G}_{\rm GUT}$ breaking. From this point of view
the situation is apparently similar to the non-unified case: the four 
 ${\mathcal G}_{\rm GUT}$ spurions can be put in one-to-one 
correspondence with the low-energy spurions $\lambda_u$,$\lambda_d$, 
$\lambda_e$, and $\lambda_\nu$. However, the smaller flavour group
does not allow the diagonalization of $\lambda_d$ and
$\lambda_e$ (which transform in the same way under ${\mathcal G}_{\rm GUT}$)
in the same basis. As a result, two additional mixing matrices 
can appear in the expressions for flavour changing rates \cite{Grinstein:2006cg}.
The hierarchical texture of the new mixing matrices is known
since they reduce to the identity matrix in the limit 
$\lambda_e^T = \lambda_d$. Taking into account this fact, and
analysing the structure of the allowed higher-dimensional operators, 
a number of reasonably  firm phenomenological consequences  
can be deduced~\cite{Grinstein:2006cg}: 
\begin{itemize}
\item 
There is a well defined limit in which the standard MFV scenario 
for the quark  sector is  fully recovered: $M_\nu \ll 10^{12}$ GeV
and small $\tan \beta$ (in a two-Higgs doublet case). 
For $M_\nu \sim  10^{12}$ GeV and small $\tan \beta$, deviations from the standard MFV pattern 
can be expected in rare $K$ decays but  not in $B$ physics.\footnote{~The 
conclusion that $K$ 
decays are the most sensitive probes of possible deviations from the  
strict MFV ansatz follows from the strong suppression of 
the $s \to d$ short-distance amplitude in the SM [$V_{td}V_{ts}^* =\cO(10^{-4})$],
and goes beyond the hypothesis of an underlying GUT. 
This is the reason why $K \to \pi \nu\bar\nu$ decays, 
which are the best probes of $s \to d$ $\Delta F=1$ short-distance amplitudes \cite{RareK},
play a key role in any extension of the SM containing non-minimal sources 
of flavour symmetry breaking, as confirmed by recent analyses performed  
in the framework of the Little Higgs model with $T$ parity~\cite{LH},
and in the MSSM with non-minimal $A$ terms~\cite{KMSSM}.}
Ignoring fine-tuned 
scenarios, $M_\nu \gg  10^{12}$~GeV is excluded by the present constraints 
on quark FCNC transitions. Independently from the value of $M_\nu$, 
deviations from the standard MFV pattern can appear both in $K$ and in $B$ physics
for $\tan\beta \gsim m_t/m_b$ (see the next section). 
\item 
Contrary to the non-GUT MFV framework,  
the rate for $\mu \to e \gamma$ (and other LFV decays) cannot be 
arbitrarily suppressed by lowering the  average mass $M_\nu$ of the heavy  $\nu_R$. 
This fact can easily be understood by looking at the flavour structure 
of the relevant effective couplings, which now assume the following form:
\begin{equation}
\label{eq:MFVgut}
(\Delta^{\ell}_{\rm LR} )_{\rm MFV-GUT} = ~ c_{1}~\lambda_e
\lambda_\nu^\dagger\lambda_\nu ~+~c_{2}~\lambda_u
\lambda_u^\dagger\lambda_e ~+~c_{3}~\lambda_u
\lambda_u^\dagger\lambda_d^T\, +~\ldots
\end{equation}
In addition to the terms involving $\lambda_\nu\sim \sqrt{M_\nu}$ already 
present in the non-unified case, the GUT group allows also $M_\nu$-independent 
terms involving the quark Yukawa couplings. The latter become competitive 
for $M_\nu \lsim 10^{12}$ GeV and their contribution is such that for 
$\Lambda \lsim 10$ TeV  the  $\mu \to e \gamma$ rate is above 
$10^{-13}$ (i.e.~within the reach of  MEG~\cite{MEG}).
\item 
Improved experimental information on $\tau \to \mu \gamma$ and 
$\tau \to e \gamma$ are a now a key tool: the best observables to 
discriminate the relative size of the MLFV contributions with respect 
to the GUT-MFV ones. In particular, if the quark-induced terms turn out 
to be dominant, the  $\mathcal{B}(\tau\to\mu\gamma)/\mathcal{B}(\mu\to e\gamma)$
ratio could reach values of $\cO(10^{-4})$, allowing $\tau\to\mu\gamma$ 
to be just below the present exclusion bounds. 
\end{itemize}    

\section{The large $\tan\beta$ scenario}

The conclusions discussed in the previous section 
are very general and holds in most GUT theories. 
The large $\tan\beta$ regime represents a more  
specific corner of GUT models, which is particularly  
interesting for flavour physics. 
Tan$\beta=v_u/v_d$, denotes the ratio 
of the two Higgs vacuum expectation values, 
which in many extensions of the SM are coupled 
separately to up- and down-type quarks (consistently 
with the MFV hypothesis). 
This parameter controls the overall normalization 
of the Yukawa couplings. The regime of large $\tan\beta$ 
[$\tan\beta = \cO(m_t/m_b)$] has an intrinsic theoretical 
interest since it allows the unification of 
top and bottom Yukawa couplings, as predicted 
for instance in SO$(10)$.

%

Since the $b$-quark Yukawa coupling become $\cO(1)$, 
the large $\tan\beta$ regime is particularly interesting 
for $B$ physics, even in absence of deviations from the 
MFV hypothesis. One of the most clear phenomenological 
consequences is the suppression of the $B \to \ell \nu$ decay
rate with respect to its SM expectation \cite{Hou}.
Potentially measurable effects are expected also in
$B\to X_s \gamma$, $\Delta M_{B_s}$ and, especially, 
in the helicity-suppressed FCNC decays 
$B_{s,d}\to \ell^+\ell^-$. The recent experimental 
evidence of $\Btaun$ at Belle \cite{Btn_Belle} and Babar \cite{DMs_CDF},
the precise $\Delta M_{B_s}$ measurement by CDF \cite{DMs_CDF},
and the constantly improving bounds on  $B_{s,d}\to \ell^+\ell^-$  
by both CDF and D0 \cite{Bmm},
make this scenario particularly interesting and timing from the phenomenological 
point if view.

\begin{figure}[t]
\begin{center}
\includegraphics[width=70mm,angle=-90]{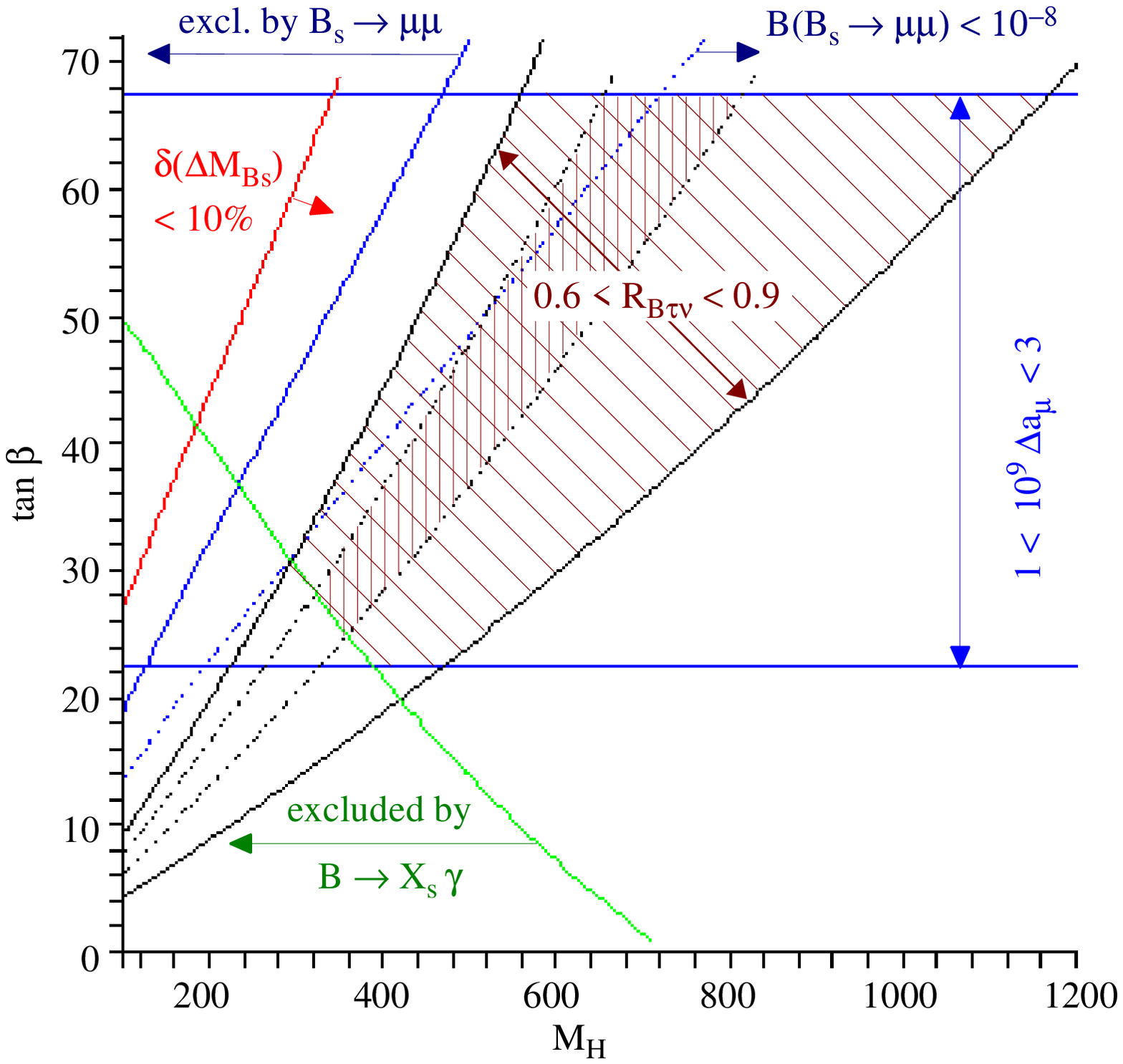}
\hskip 0.1 cm
\includegraphics[width=70mm,angle=-90]{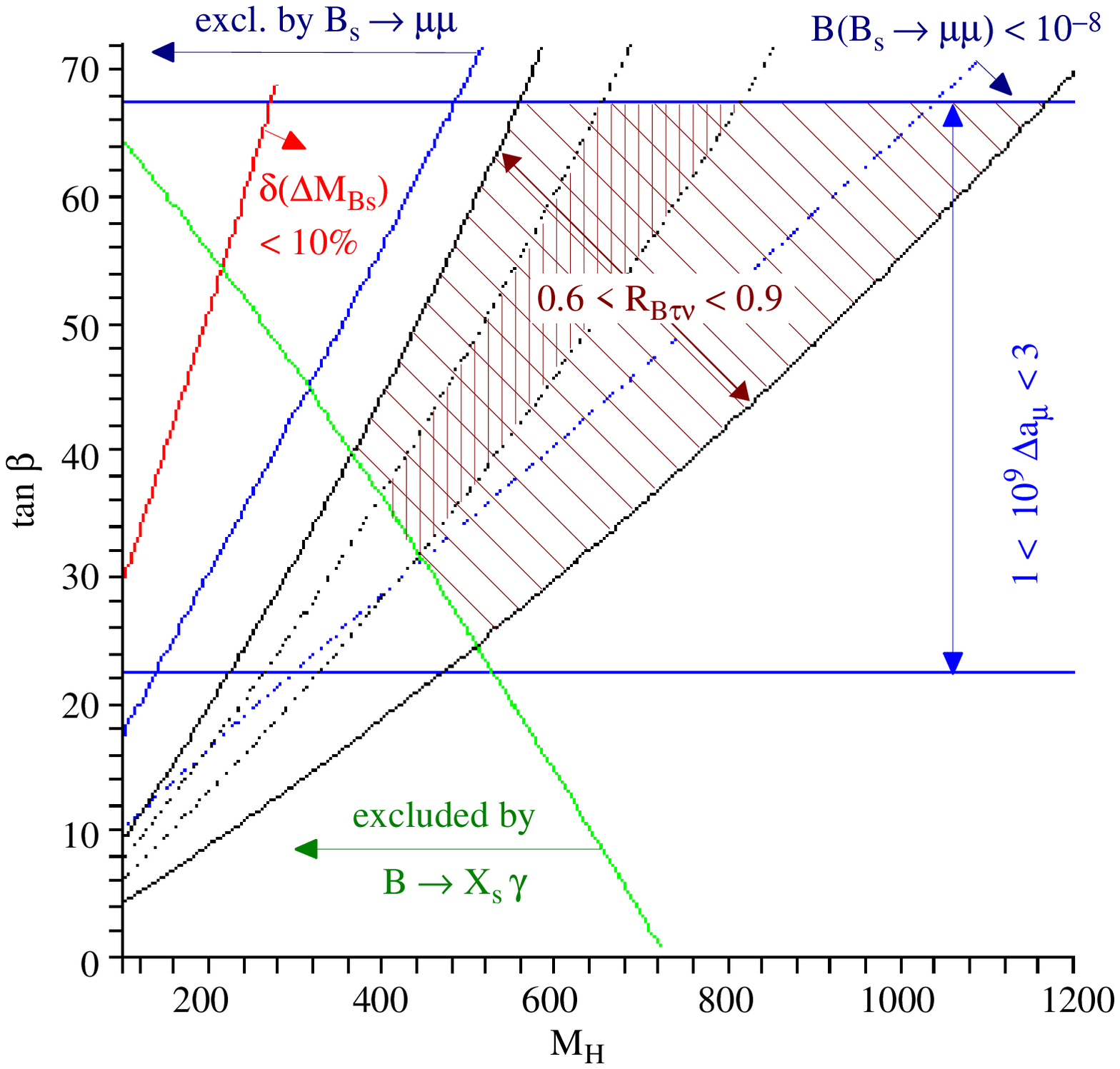}
\caption{Correlations in the $M_H$--$\tan\beta$ plane within the MSSM 
for heavy squarks ($\mu = \msq = 2 \msl = 3 M_2 \approx 1$~TeV, 
$A_U=-/\!+1$~TeV in the left/right plot) \cite{noi}. 
$R_{B\tau\nu} = \BR(\Btaun)/\BR^{\rm SM}(\Btaun)$. \label{fig:tanb}}
\end{center}
\end{figure}

Within the EFT approach where all the 
high degrees of freedom --but for the Higgs fields-- 
are integrated out \cite{MFV}, we cannot establish well-defined 
correlations among these observables. However, the scenario
becomes quite predictive within a more ambitious EFT: the MSSM
with MFV. As recently shown in Ref.~\cite{noi,vives}, in the MFV-MSSM
with large $\tan\beta$ and heavy squarks, interesting correlations
can be established among the $B$ observables mentioned above 
and two flavour-conserving observables: the anomalous magnetic moment 
of the muon and the lower limit on the lightest Higgs boson mass.
An illustration of these correlations is shown in Fig.~\ref{fig:tanb}.

Present data are far from having established a clear evidence 
for such scenario (as for any deviation from the SM). Nonetheless, 
it is interesting to note that this scenario can naturally solve 
the long-standing $(g-2)_\mu$ anomaly and explain in a natural 
why the lightest Higgs boson has not been observed yet.
Moreover, it predicts visible deviations from the SM in 
$\BR(\Btaun)$ (most likely a suppression, of at least $10\%$) 
and $B_{s,d}\to \ell^+\ell^-$ (most likely a enhancement, 
up to a factor of 10), which could possibly be revealed 
in the near future. Finally,  the parameter space which leads to these 
interesting effects can also naturally explain why 
$\BR(B\to X_s \gamma)$ and $\Delta M_{B_s}$ are
in good agreement with the SM expectations \cite{noi}.

The observables $\BR(\Btaun)$,  $\BR(B_{s,d}\to \ell^+\ell^-)$
and $(g-2)_\mu$ can be considered as the most promising 
low-energy probes of the MSSM scenario with heavy squarks and large $\tan\beta$. 
Nonetheless, interesting consequences of this scenario could 
possibly be identified also in other observables.
In particular, as pointed out in \cite{kl2}, if the slepton sector 
contains sizable sources of LFV, we could even hope to observe 
violations of lepton universality in the 
$\BR(P\to \ell \nu)/\BR(P \to \ell^\prime \nu)$
ratios. Deviations from the SM can be $\cO(1\%)$ 
in $\BR(K\to e \nu)/\BR(K \to \mu \nu)$ \cite{kl2}, 
and  can reach $\cO(1)$ and $\cO(10^3)$ in  
$\BR(B\to \mu \nu)/\BR(B \to \tau \nu)$ and 
$\BR(B\to  e \nu)/\BR(B \to \tau \nu)$, respectively~\cite{noi}.

\section{Conclusions} 
Rare decays of quarks and leptons provide a unique 
opportunity to shed more light on the underlying 
mechanism of flavour mixing. The discovery of neutrino 
oscillations, which has opened a new era in flavour 
physics, gives us more hopes in this respect. 
As we have shown by means of a general EFT approach 
to physics beyond the SM, under rather general hypothesis
(quark-lepton unification and new physics in the TeV range)
neutrino oscillations imply the existence of interesting 
non-standard effects also in rare decays of charged leptons and, 
possibly, in a few rare $B$ and $K$ decay observables. 
 
The most solid and exciting expectation is a $\mu \to e\gamma$ 
branching ratio exceeding $10^{-13}$, i.e.~within the reach of 
the MEG experiment. In more specific scenarios, we could also observe 
sizable non-standard effects in rare FCNC  
$\tau$ and $K$ decays and --particularly 
in the large $\tan\beta$ regime of the MSSM--  in the purely leptonic 
decays of both charged and neutral $B$ mesons.

\newpage

\section*{Acknowledgments} 
It is a pleasure to thank
Vincenzo Cirigliano, Ben Grinsten, Paride Paradisi,
Valentina Porretti, and Mark Wise, for the enjoyable 
collaborations on which this talk is largely based. 
I also wish to thank organizers of HQL2006 
for the invitation to this interesting conference. 
This work has been supported in part by the EU Contract No.
MRTN-CT-2006-035482, ``FLAVIAnet''.

\end{document}